\documentclass[aps,prb,superscriptaddress,twocolumn,showpacs]{revtex4-1}
\usepackage[english]{babel}
\usepackage{graphicx}
\makeindex

\usepackage{amsmath,amssymb}
\usepackage[utf8]{inputenc}
\usepackage{natbib}
\usepackage{units}
\usepackage{hyperref}
\usepackage{color,soul}
\hypersetup{colorlinks=true,citecolor={blue},linkcolor={blue},urlcolor={blue}}
\usepackage{notes2bib}
\usepackage{balance}
\usepackage{yfonts}
\usepackage{float}

\newif\iffig\figfalse

\begin{document}
\title{Switching waves in multi-level incoherently driven polariton condensates}

\date{\today}

\author{H. Sigurdsson}
\email[correspondence address:~]{helgi001@e.ntu.edu.sg}
\affiliation{Division of Physics and Applied Physics, Nanyang Technological University 637371, Singapore}
\affiliation{Science Institute, University of Iceland, Dunhagi-3, IS-107 Reykjavik, Iceland}

\author{I. A. Shelykh}
\affiliation{Division of Physics and Applied Physics, Nanyang Technological University 637371, Singapore}
\affiliation{Science Institute, University of Iceland, Dunhagi-3, IS-107 Reykjavik, Iceland}
\affiliation{ITMO University, St. Petersburg, 197101, Russia}

\author{T. C. H. Liew}
\affiliation{Division of Physics and Applied Physics, Nanyang Technological University 637371, Singapore}

\begin{abstract}
We show theoretically that an open-dissipative polariton condensate confined within a trapping potential and driven by an incoherent pumping scheme gives rise to bistability between odd and even modes of the potential. Switching from one state to the other can be controlled via incoherent pulsing which becomes an important step towards construction of low-powered opto-electronic devices. The origin of the effect comes from modulational instability between odd and even states of the trapping potential governed by the nonlinear polariton-polariton interactions.
\end{abstract}

\maketitle

\section{Introduction}
Switching waves arising in bistable light-matter systems are of fundamental importance for future opto-electronic devices that intertwine both the characteristics of electrons and photons. Such waves can be thought of as moving domain walls (wave crests) where one state succumbs to its bistable counterpart. In general, light-matter systems owe their bistability to their optical non-linear nature. An example of such a system is the optical microcavity~\cite{microcavities} where strong coupling between excitons living in embedded quantum wells, and cavity photons gives rise to the light exciton-polariton quasiparticle (or simply, polariton)~\cite{weisbuch_1992}. In the past decade, a great amount of research has been dedicated to these light-matter bosons. Among the most interesting properties of the polariton is its ability to condense into a macroscopically occupied coherent state~\cite{kasprzak_2006}, essentially a Bose-Einstein condensate, at temperatures much higher than conventional atomic condensates~\cite{christopoulos_2007} due to its light effective mass.

It has been shown that exciton-polariton systems give rise to bistability under resonant pumping schemes~\cite{gippius_2004, baas_2004, whittaker_2005} with applications in photonic logic gates~\cite{amo_2010, adrados_2011, ortega_2013},  switches and memory elements~\cite{cerna_2013, cancellieri_2014, grosso_2014}, control of polariton superfluids~\cite{giorgi_2012}, and optical diodes~\cite{ortega2_2013}. Under resonant excitation a number of works focused on the parametric instability between polaritons~\cite{savvidis_2000, ciuti_2000, ciuti_2001, saba_2001, tartakovskii_2002}, where polaritons in one particular energy-momentum state scatter in pairs to "signal" and "idler" states conserving energy and momentum. Theoretically, a bistability of the system was shown to be associated with the process~\cite{whittaker_2005}, corresponding to the stability of states under the same conditions where the scattering could be activated or not. Bistability resulting from parametric interaction was further shown to allow for the formation of solitons~\cite{egorov_2010, egorov_2013, egorov_2014}. 

To lift the requirement of a resonant laser, a few works have considered mechanisms to generate bistability under non-resonant or incoherent excitation, but typically required specially designed structures~\cite{savenko_2014}. Theoretical work showed that parametric instability and an associated bistability could be arranged in sub-wavelength grating microcavities~\cite{kyriienko_2014}. Recently, the existance of spin-based bistability induced by different $XY$-polarization lifetimes was shown experimentally~\cite{ohadi_2015}. It has also been shown that modulational instabilities can lead to new states that co-exist with stable states~\cite{liew_2015}.

In this work we study a mechanism of bistability, appearing when polaritons are incoherently excited in the presence of a trapping potential. Different modes of the trapping potential can be preferentially excited via the spatial patterning of the incoherent excitation. Non-linear coupling between the modes generated by polariton-polariton interactions then leads to a modulational instability. Remarkably, a bistability is associated with this process and switching between different bistable states can be achieved with incoherent pulses, even when only the first two eigenmodes are considered. In extended 2D systems, switching waves that determine the boundary between spatially separated domains can be excited.

The article is structured as follows: In Sec.~\ref{sec.model} we introduce the model describing the coherent polariton condensate confined within a trapping potential. In Sec.~\ref{sec.2mode} we present results where we limit our analysis to only the first two eigenstates of the harmonic potential. In Sec.~\ref{sec.3mode} we present results considering only the first three eigenstates of the harmonic potential. In Sec.~\ref{sec.cGP} we present results considering the full theoretical model containing all eigenstates of the harmonic potential. Here we demonstrate a switching wave signal traveling in a 2D planar system. In Sec.~\ref{sec.IQW} we demonstrate analogous bistability in another type of trapping geometry, the infinite quantum well.
 
\section{Theoretical model} \label{sec.model}
We consider a microcavity with an embedded 2D quantum well subjected to a trapping potential~\cite{balili_2006, Balili_2007}. In an open-dissipative system where the incoherent pumping is balanced against the polariton lifetime and an exciton reservoir decay mechanism~\cite{wouters_2007}, the true solutions of the interacting condensate confined in a trapping potential are in fact non-trivial~\cite{berman_2008}. The trapping potential eigensolutions (modes) in the 1D Schrödinger equation are termed $\psi_n(x)$ with energies $E_n$ and form a complete basis such that we can always write our condensate order parameter in this basis. Assuming that the eigenenergies $E_n$ are different, then in order for the condensate to become stationary, populations in different modes are blueshifted such to become degenerate, allowing polaritons freely to 'flow' from one mode to another.
\begin{figure}[!t]
  \centering
		\includegraphics[width=0.38 \textwidth]{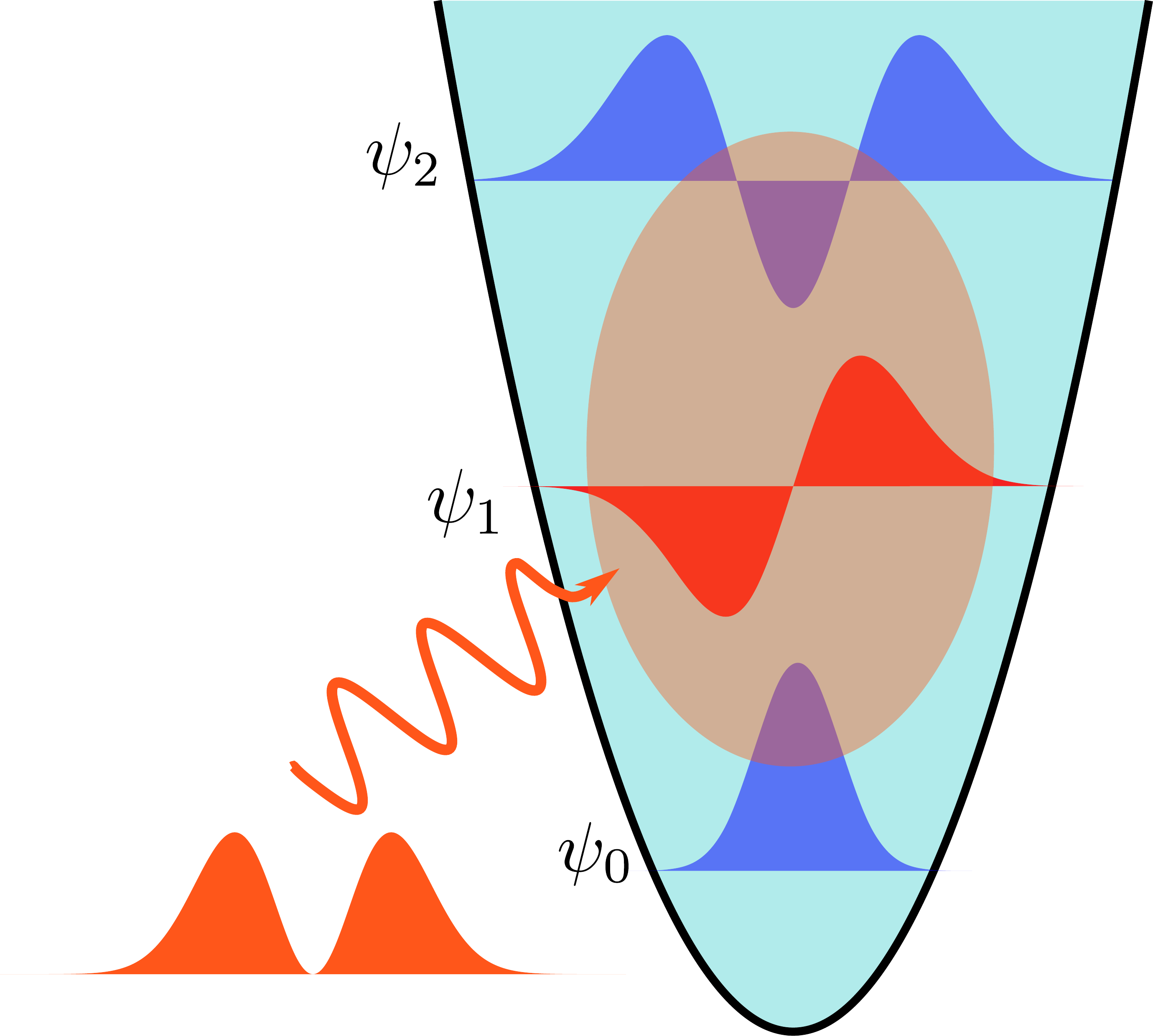}
		\caption{(Color online) Example schematic showing the case of a harmonic potential, $V(x) = m\omega^2 x^2/2$ and its first three states. The incoherent pump $P(x)$ (orange) is chosen to have maximum overlap with one state but will also have some finite overlap with higher states thus creating a condensate of coupled modes.}
\label{fig.sketch}
\end{figure}

To account for the aforementioned effects, the spatial dynamics of interacting polaritons can be described by the complex Gross-Pitaevskii (cGP) equation~\cite{keeling_2008},
\begin{align} \notag
i \hbar \frac{d \Psi}{d t} & = \bigg[-\frac{\hbar^2}{2m} \frac{\partial^2}{\partial x^2} + \alpha |\Psi|^2 + V(x) \dots \\  \label{eq.ham}
& + i  \left(  P(x) - \frac{\hbar}{2 \tau} - R |\Psi|^2 \right) \bigg]\Psi.
\end{align}
Here $\Psi$ is the polariton condensate order parameter, $m$ is the polariton effective mass, $\alpha$ is the interaction strength, $\tau$ is the polariton lifetime, $R$ is the condensate saturation (also known as reservoir depletion rate), $P(x)$ is the spatial profile of the incoherent pump intensity, and $V(x)$ is the trapping potential.

The condensate order parameter can be written,
\begin{equation} \label{eq.ordP}
\Psi(x,t) = \sum_{n=0}^\infty A_n(t) \psi_n(x).
\end{equation}
The time evolution of the cGP-equation is governed by the modes amplitudes $A_n(t)$. In order to favor the generation of polaritons in the first excited state the pump takes the following profile,
\begin{equation} \label{eq.pump}
P(x) = P_0 |\psi_1(x)|^2
\end{equation}
where $P_0$ denotes the intensity of the pump. While the pump is chosen so as to have a maximum overlap for the $A_1$ mode making it condense first, it understandably has non-zero overlap with other modes (see Fig.~\ref{fig.sketch}). This results in an interplay between the gain and the decay of all the modes. Substituting Eqs.~\ref{eq.ordP}-\ref{eq.pump} into Eq.~\ref{eq.ham} and integrating over the $n$-th state, $\psi_n$, we come to the dynamical equation for the condensate modes,
\begin{align} \notag
i \hbar \frac{dA_n}{dt} = & \left( E_n - \frac{i \hbar}{2 \tau} \right) A_n + i P_0 \sum_ m M_{n11m} A_m \dots \\ \label{eq.ham2}
& +  (\alpha - i R) \sum_{jkl} M_{njkl} A_j^* A_k A_l. 
\end{align}
Here $M_{ijkl}$ are transition matrix elements that obey,
\begin{equation}
M_{ijkl} = \int_{-\infty}^\infty \psi_i \psi_j \psi_k \psi_l \ dx.
\end{equation}

\section{Two-Mode Model} \label{sec.2mode}
In order to test our theory for the most simple case, we investigate whether bistability can exist considering only scattering between the first two states of the 1D harmonic oscillator, $\psi_0$, and $\psi_1$, assuming that modulational instability takes place. Modulational instability was also predicted for non-resonantly excited polariton condensates where the condensate has a strong back-action effect on an incoherent reservoir of hot exciton states~\cite{smirnov_2014, bobrovska_2014}. We are not in this regime as the complex Gross-Pitaevskii approach~\cite{keeling_2008}, corresponding to Eq.~\ref{eq.ham}, applies to the case where the reservoir has a faster timescale than polariton dynamics and can be treated as independent. 

\begin{figure}[!t]
  \centering
		\includegraphics[width=0.48 \textwidth]{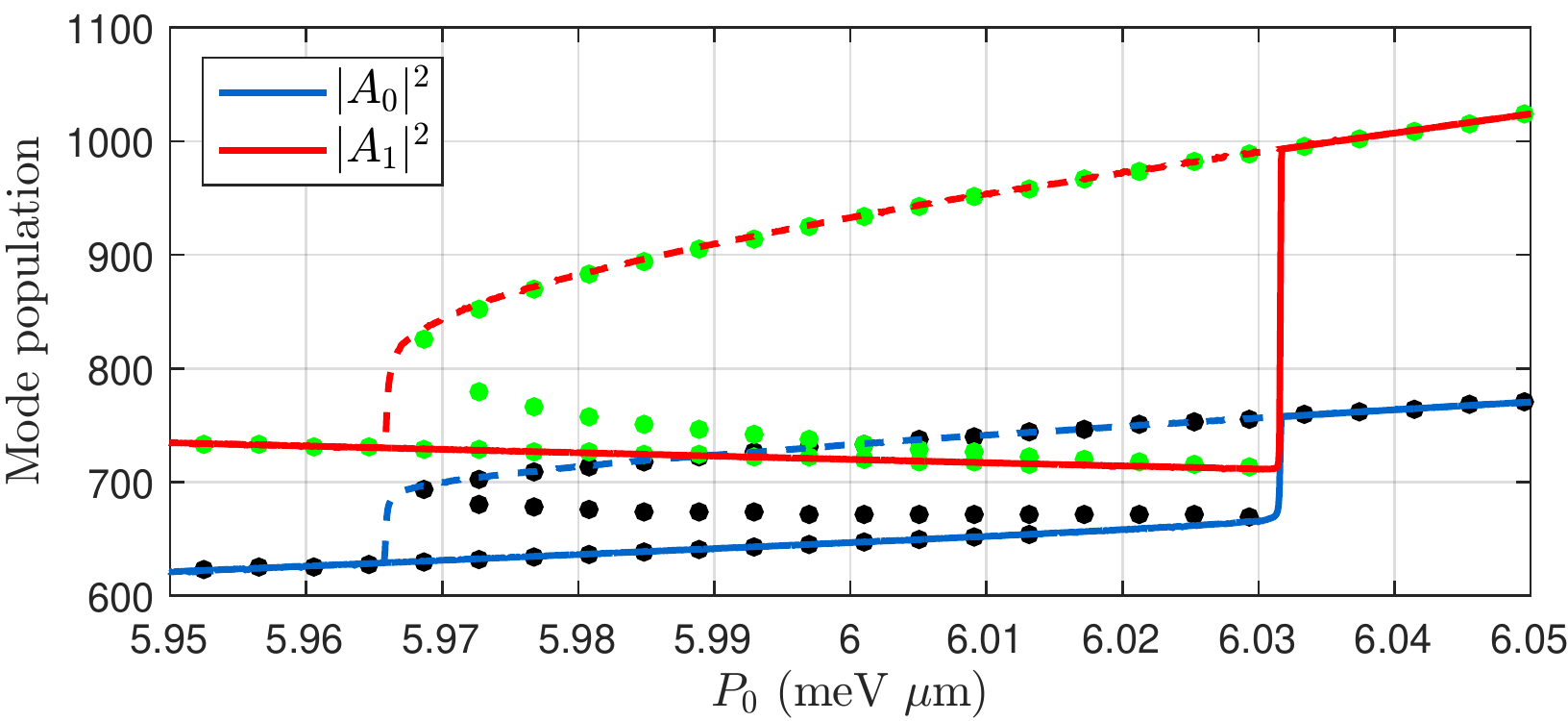}
		\caption{(Color online) Whole (dashed) blue and red lines show the time resolved density from Eqs.~\ref{eq.mode21}-\ref{eq.mode22} when the pump is slowly switched on (off). Calculated solutions from the analysis leading up to Eq.~\ref{eq.cube} (green and black dots). Parameters: $m = 3 \cdot 10^{-5} m_0$, $\hbar \omega = 51.5$ $\mu$eV, $\alpha = 2.4$ $\mu$eV $\mu$m, $\tau = 2$ ps and $R = 0.3\alpha$.}
\label{fig.2mode}
\end{figure}

The condensate order parameter can be written in the limited basis of the first two harmonic oscillator states,
\begin{equation} \label{eq.2mode}
\Psi(x,t) = A_0(t) \psi_0(x) + A_1(t) \psi_1(x).
\end{equation}
A complete analysis is given in appendix~\ref{sec.app1}. We begin by adiabatically switching on the pump. A condensation threshold is reached for the $A_1$ mode which, in the low density regime, satisfies $P_0 = \hbar/2 \tau M_{1111}$. At higher pumping powers we find that indeed there exists a bistable regime (see Fig.~\ref{fig.2mode}) where switching from one solution to the other can take place by pulsing the system incoherently. The blue and red solid (dashed) lines show results of slowly increasing (decreasing) the pump. The green and black dots are semi-analytical results given by Eq.~\ref{eq.cube} in appendix~\ref{sec.app1}. One can see that the central branches of the hysteresis curve are resolved from the Eq.~\ref{eq.cube} whereas numerically time resolved results only show the bottom and the top branches.

In the case of a noninteracting condensate where $\alpha = 0$ and setting $\tau \to \infty$ we find stationary solutions corresponding to the case where only one mode stays populated (see appendix~\ref{sec.app1}). Numerically we observe no solutions where both modes are populated at the same time. Indeed, from Eqs.~\ref{eq.mode21}-\ref{eq.mode22} it is not directly evident whether such a solution exists or not. Thus, in theory, it remains unproven whether bistability exists in the absence of interactions. However, in Sec.~\ref{sec.cGP}[a] we will demonstrate that the full noninteracting cGP-model yields population higher modes.

\section{Three-Mode Model} \label{sec.3mode}
\begin{figure}[!t]
  \centering
		\includegraphics[width=0.48 \textwidth]{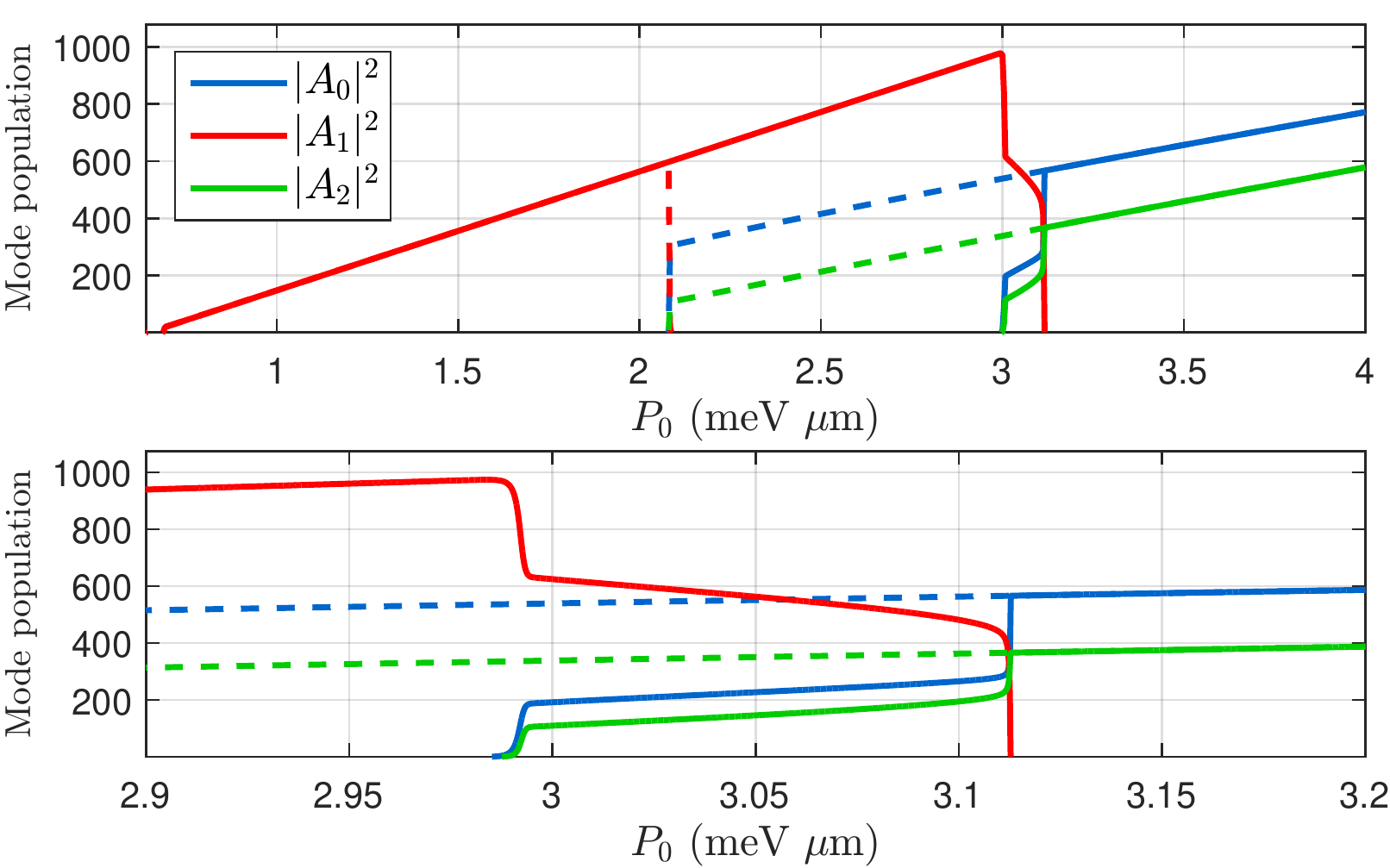}
		\caption{(Color online) Results of the three-mode model under adiabatic incoherent pumping. (a) Population of the modes shown when the pump is slowly switched on (solid lines) and then slowly turned off (dashed lines). At a certain pump threshold there is a transition from $A_1$ into a condensate composed of all three modes. At higher pump values only $A_0$ and $A_2$ stay populated. One can see that a bistable interval exists at $2.1$ meV $\mu$m $ \lesssim P_0 \lesssim 3.1$ meV $\mu$m.  (b) A close up of the transition from frame (a). Parameters: $m = 3 \cdot 10^{-5} m_0$, $\hbar \omega = 34$ $\mu$eV, $\alpha = 2.4$ $\mu$eV $\mu$m, $\tau = 15$ ps and $R = \alpha$.}
\label{fig.res1}
\end{figure}

\begin{figure}[!hbtp]
  \centering
		\includegraphics[width=0.48 \textwidth]{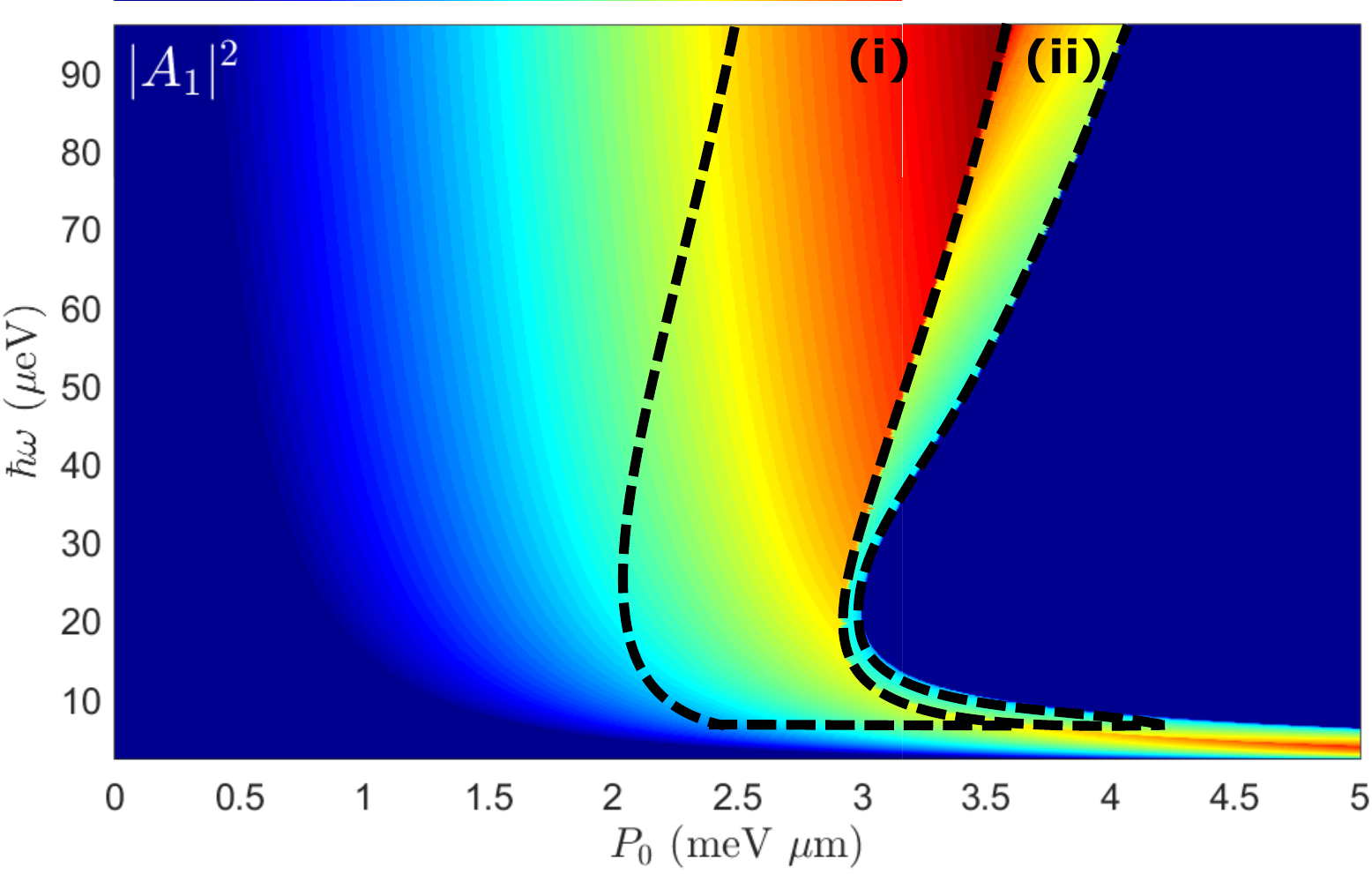}
		\caption{(Color online) Population of the second mode $A_1$ shown for varying oscillator frequency and pump intensity. Dashed lines enclose the two different bistable regions corresponding to those demonstrated in Fig.~\ref{fig.res1}. Same parameters are used as in Fig.~\ref{fig.res1}. }
\label{fig.res2}
\end{figure}

We now extend our model to the first three eigenmodes of the HO where the order parameter is written, 
\begin{equation} \label{eq.2mode}
\Psi(x,t) = A_0(t) \psi_0(x) + A_1(t) \psi_1(x) + A_2(t) \psi_2(x).
\end{equation}
Solving the three mode model numerically (see Eqs.\ref{eq.3mode1}-\ref{eq.3mode3}) we find that bistable regions can indeed exist through a wide range of oscillator frequencies $\omega$ and pump intensities $P_0$ (see Figs.~\ref{fig.res1} and~\ref{fig.res2}). Results displayed in Fig.~\ref{fig.res1} show the change in population of the three modes as the non-resonant pump is slowly switched on (whole lines) and then slowly switched off (dashed lines). First, a condensation threshold is reached for the $A_1$ mode which in the low density regime satisfies $P_0 = \hbar/2 \tau M_{1111} \simeq 0.65$ meV $\mu$m. At higher pumping values there is an abrupt drop in the strength of the $A_1$ mode as scattering and gain of the $A_0$ and $A_2$ modes overcome their decay. Just as with the two-mode model, the three modes are degenerate and co-exist at the same energy at all times when they are populated. As the pump is increased further, the $A_1$ mode is completely quenched and only mode $A_0$ and $A_2$ stay populated.

When the pump is slowly decreased (dashed lines in Fig.~\ref{fig.res1}), one can clearly see that modes $A_0$ and $A_2$ remain supported over an interval of pump values where previously only the $A_1$ mode was supported. A phase diagram showing the bistable region across different values of oscillator energies $\hbar \omega$ and pump intensities $P_0$ is displayed in Fig.~\ref{fig.res2}. The black dashed lines indicate the two different bistable areas. The larger one, (i), shows where only $A_1$ exists but can be excited to a solution where $A_0$ and $A_2$ only exist. The smaller one, (ii), shows where all modes are populated but can also be excited to an $A_0$ and $A_2$ only solution.

Analogous to our results in Sec.~\ref{sec.2mode}, numerically we observe no population in the $A_0$ and $A_2$ modes when interactions are absent.

\section{Full cGP-formalism} \label{sec.cGP}
So far simplified models have been considered using only the first two- and three eigenmodes of the HO. These have demonstrated two artifacts of the trapped incoherently generated polariton condensate. Namely, a bistable interval between the mode populations (Sec.~\ref{sec.2mode}-\ref{sec.3mode}), and a sudden transition from an odd solution into an even one (Sec.~\ref{sec.3mode}).

Realistic polariton systems however would contain many spatial modes, corresponding to a full model of the polariton spatial dynamics according to Eq.~\ref{eq.ham}. Here, we solve the Eq.~\ref{eq.ham} directly by propagating it stepwise in time and demonstrate that both of these artifacts from previous sections are still clearly present.

\subsection{1D HO System}
\begin{figure}[!t]
  \centering
		\includegraphics[width=0.48 \textwidth]{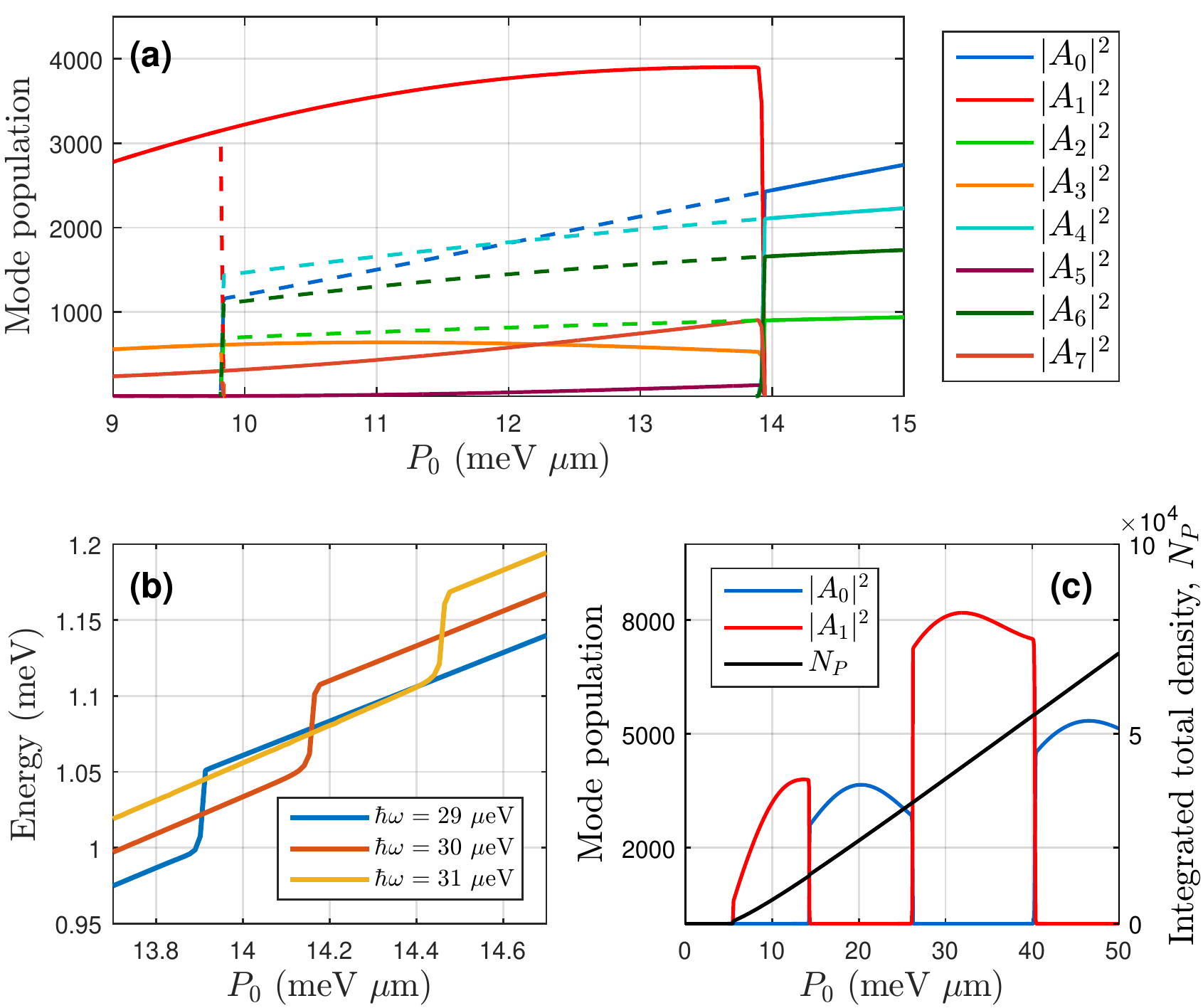}
		\caption{(Color online) (a) Populations of the first eight harmonic modes using the full cGP formalism. A bistable interval is observed at $9.8$ meV $\mu$m $ \lesssim P_0 \lesssim 13.9$ meV $\mu$m when pump is increased (solid lines) and then decreased (dashed lines) associated with odd modes transitioning into even ones and back. (b) Condensate energy during the forward transition for different potential strengths. (c) Increasing the pump strength reveals that the transition from odd to even modes takes place periodically with increasing pump power and does not affect the total integrated density $N_P$ (black line). Note that frames (b-c) are only shown for increasing pump strength. Parameters were set to: $m = 3 \cdot 10^{-5} m_0$, $\tau = 2$ ps, $\alpha = 2.4$ $\mu$eV $\mu$m and $R = 0.3 \alpha$. In frames (a,c): $\hbar \omega = 30$ $\mu$eV.}
\label{fig.res1D}
\end{figure}

Results for a 1D system are shown in Fig.~\ref{fig.res1D}, calculated for typical parameters of a GaAs based system:  $m = 3 \cdot 10^{-5} m_0$, where $m_0$ is the free electron mass, $\alpha = 2.4$ $\mu$eV $\mu$m, $\tau = 2$ ps and $R = 0.3\alpha$. As the pump intensity is slowly increased we observe population growth only in the odd modes with the fastest growth in $A_1$. As the pump intensity increases further a sudden transition takes place where the population of the odd modes ($A_1$ included) suddenly shifts to the even modes. This is analogous to the results displayed in Fig.~\ref{fig.res1} where $A_1$ succumbs to $A_0$ and $A_2$. Slowly decreasing the pump (dashed lines in Fig.~\ref{fig.res1D}[a]) reveals a bistable region before the condensate transitions back to a superposition of odd modes. Both solutions of the condensate are found to be stable within $9.8$ meV $\mu$m $ \lesssim P_0 \lesssim 13.9$ meV $\mu$m. From Fig.~\ref{fig.res1D}[a] it can be seen that the population strength of different modes does not follow a strict hierarchy, an example is the crossover of $|A_3|^2$ and $|A_7|^2$. The energy of the condensate reveals a sudden blueshift when the transition takes place (see Fig.~\ref{fig.res1D}[b]) which increases with the oscillator energy $\hbar \omega$. Interesting enough, the observed transition takes place periodically with increasing pump intensity (see Fig.~\ref{fig.res1D}[c]), and using pump spatial profiles corresponding to higher harmonic states also induces a transition from and odd to an even condensate. In Fig.~\ref{fig.res1D}[c] we see that the total integrated density $N_P$ (black line) is unaffected by these transitions. We furthermore observe phase-locking taking place between the two condensate solutions as the condensate is transitioning from one to another, corresponding to coherent transport of polaritons between odd and even states. Such phase synchronization phenomenon has been previously reported in spinor polariton condensates due to parametric scattering~\cite{walker_2011}. In the noninteracting case $(\alpha = 0)$ the transition is no longer observed for increasing pump power and the polariton population remains only in the odd modes, corresponding to the nontrivial stationary state of the pumped noninteracting condensate. This is in analogy with results from Secs.~\ref{sec.2mode}-\ref{sec.3mode} where no population could be observed in the neighboring even modes, $A_0$ and $A_2$.

\subsection{2D HO System}
\begin{figure}[!t]
  \centering
	\includegraphics[width=0.5 \textwidth]{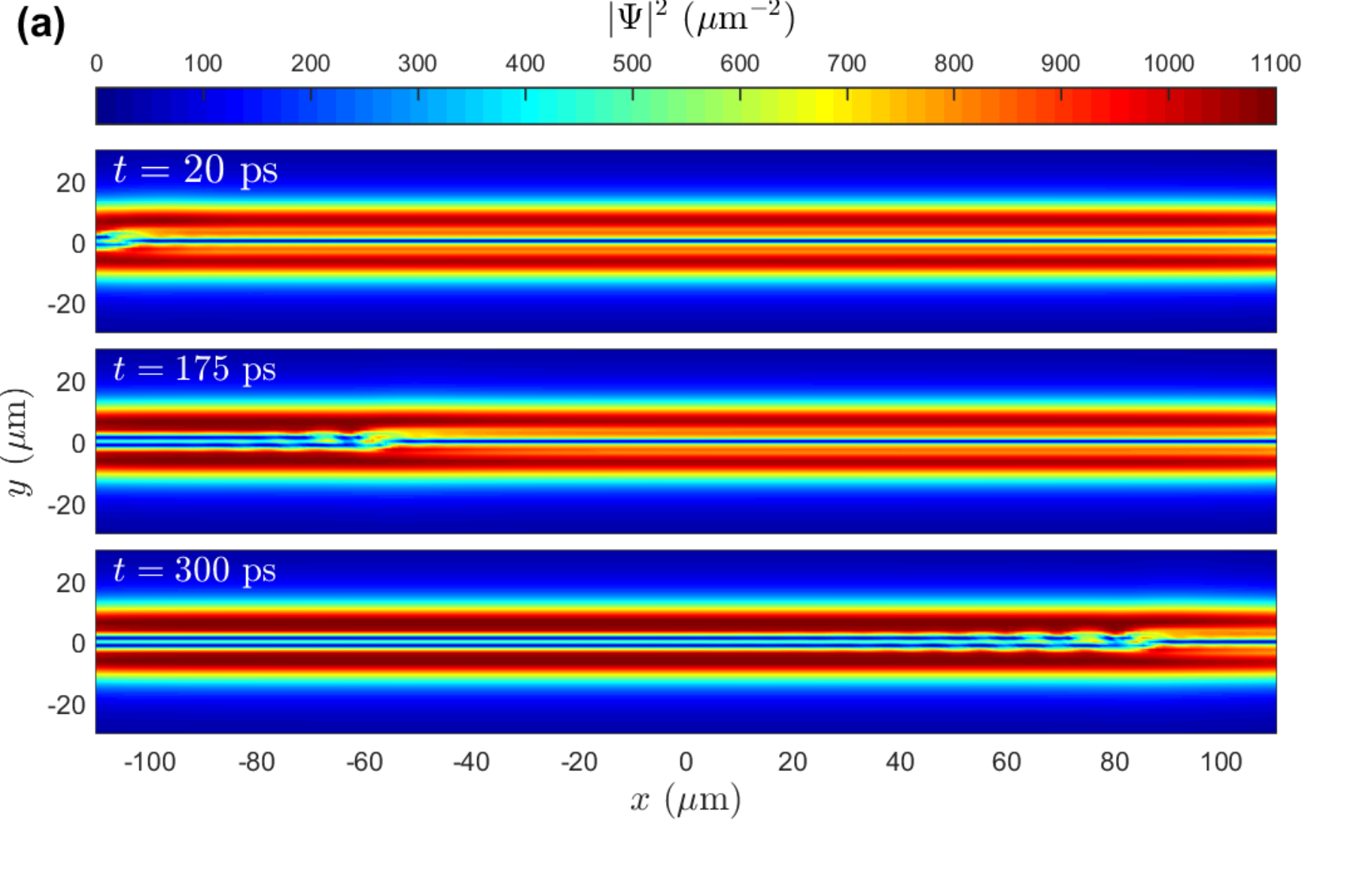} 
	\includegraphics[width=0.5 \textwidth]{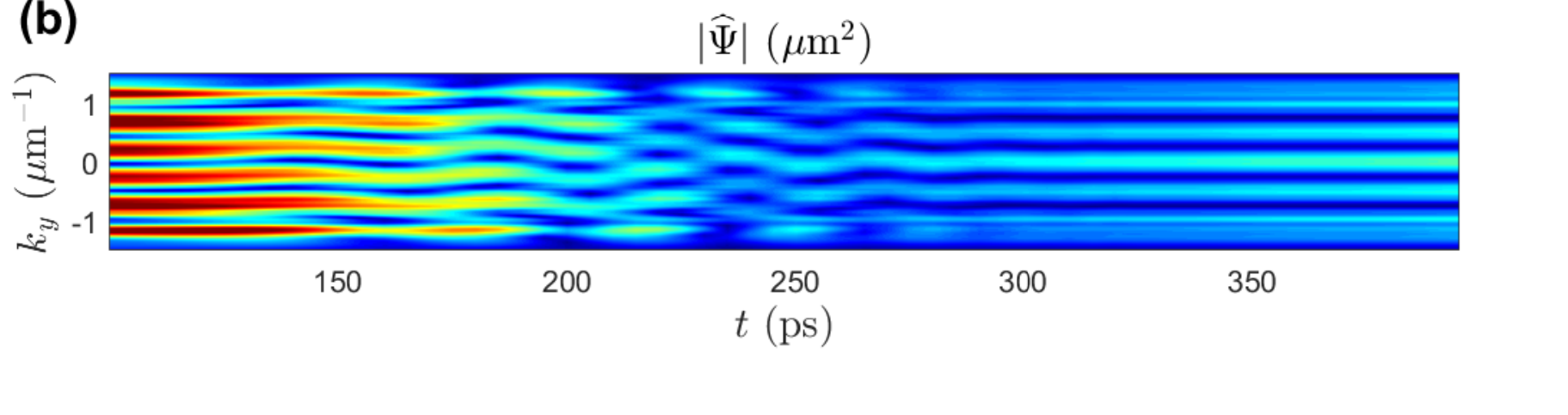} 
	\includegraphics[width=0.5 \textwidth]{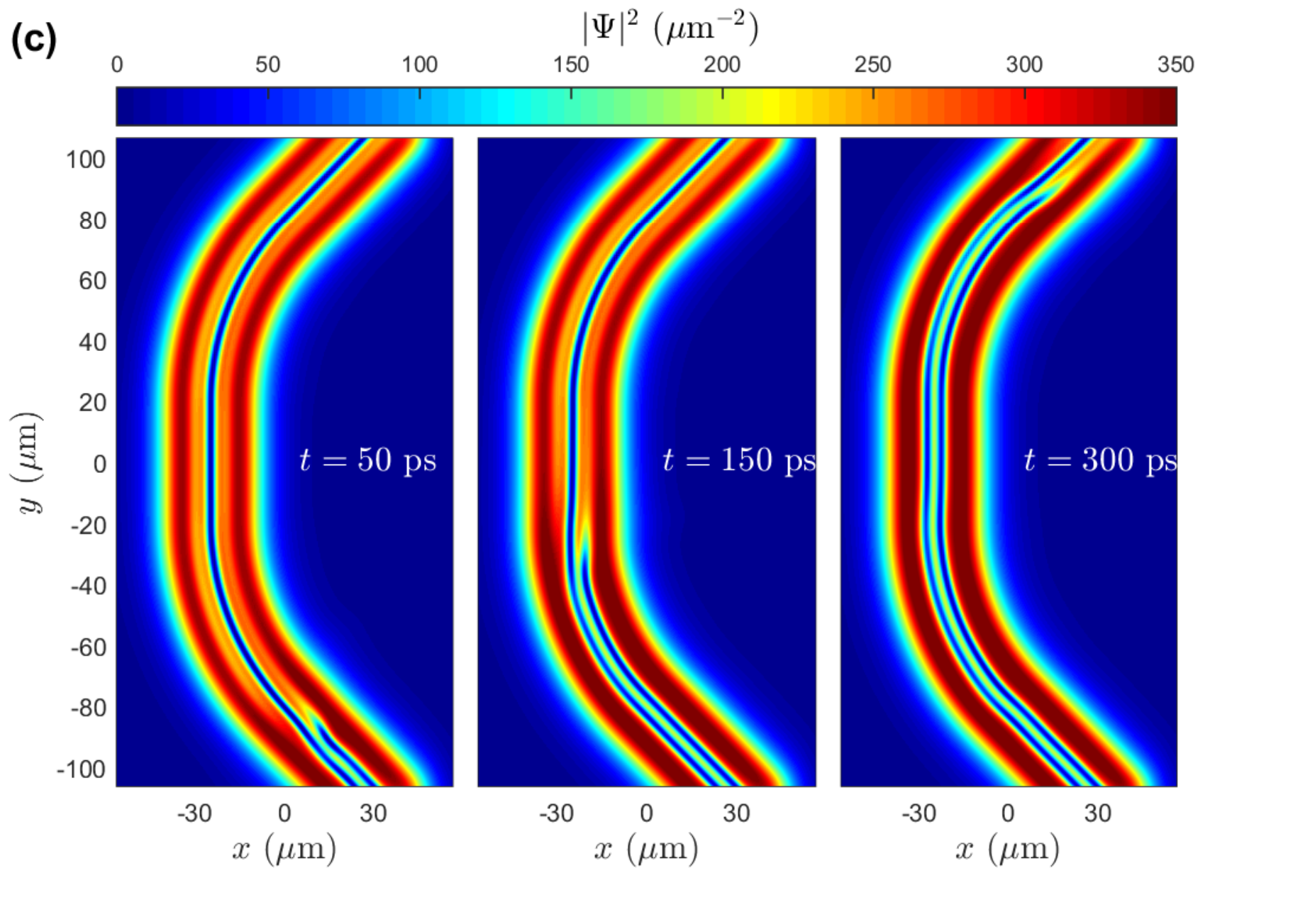} 
		\caption{(Color online) (a) Switching wave demonstrated in a 2D system with a parabolic trap of strength $\hbar \omega = 60$ $\mu$eV along the $x$-axis. An incoherent pulse is activated causing a sudden transition from a condensate composed of odd harmonic states into even states in a form of a switching wave. (b) $K$-space dynamics showing change of topology when the transition occurs. (c) A switching wave traveling along curved potential geometry where $\hbar \omega = 30$ $\mu$eV. Other parameters were set to same values used in Fig.~\ref{fig.res1D}.}
\label{fig.res3}
\end{figure}

Applying the 1D system to a 2D one, the polariton condensate can be confined along a parabolic guide. By slowly switching on the non-resonant background pump to the bistable regime, a nice steady state solution, superposed of odd modes, is formed along the guide. The condensate is then pulsed incoherently at one end resulting in a switching wave which travels along the guide (this can also be achieved by coherent injections) transitioning odd harmonic modes into even modes. In Fig.~\ref{fig.res3}(a) we see a signal traveling along a straight guide, set to $\hbar \omega = 60$ $\mu$eV, after a pulsed incoherent injection on the left end. At first the signal moves slowly, then it speeds up as the condensate start to switch rapidly into its bistable counterpart. From numerical calculations the velocity of the signal is determined to be around $\sim 1.2$ $\mu$m/ps. 

The temporal width of the pulse is in the order of a few picoseconds. Fig.~\ref{fig.res3}(b) shows the corresponding change in $K$-space occupancy, extracted from $x = 0$ in Fig.~\ref{fig.res3}(a).  In Fig.~\ref{fig.res3}(c) we demonstrate the same type of a switching wave but now in a guide with $\hbar \omega = 30$ $\mu$eV. Here the signal travels successfully around a curved guide completing a 90 degree turn. Such switching waves can be realized over a large range of potential frequencies. In fact, increasing the trap frequency $\omega$ reveals that the blueshift associated with the switching increases (Fig.~\ref{fig.res1D}[b]) and consequently increases the speed of the signal  This opens up the possibility of controlling the signal speed using different potential strengths.

\section{Infinite Quantum Well} \label{sec.IQW}
\begin{figure}[!t]
  \centering
	\includegraphics[width=0.5 \textwidth]{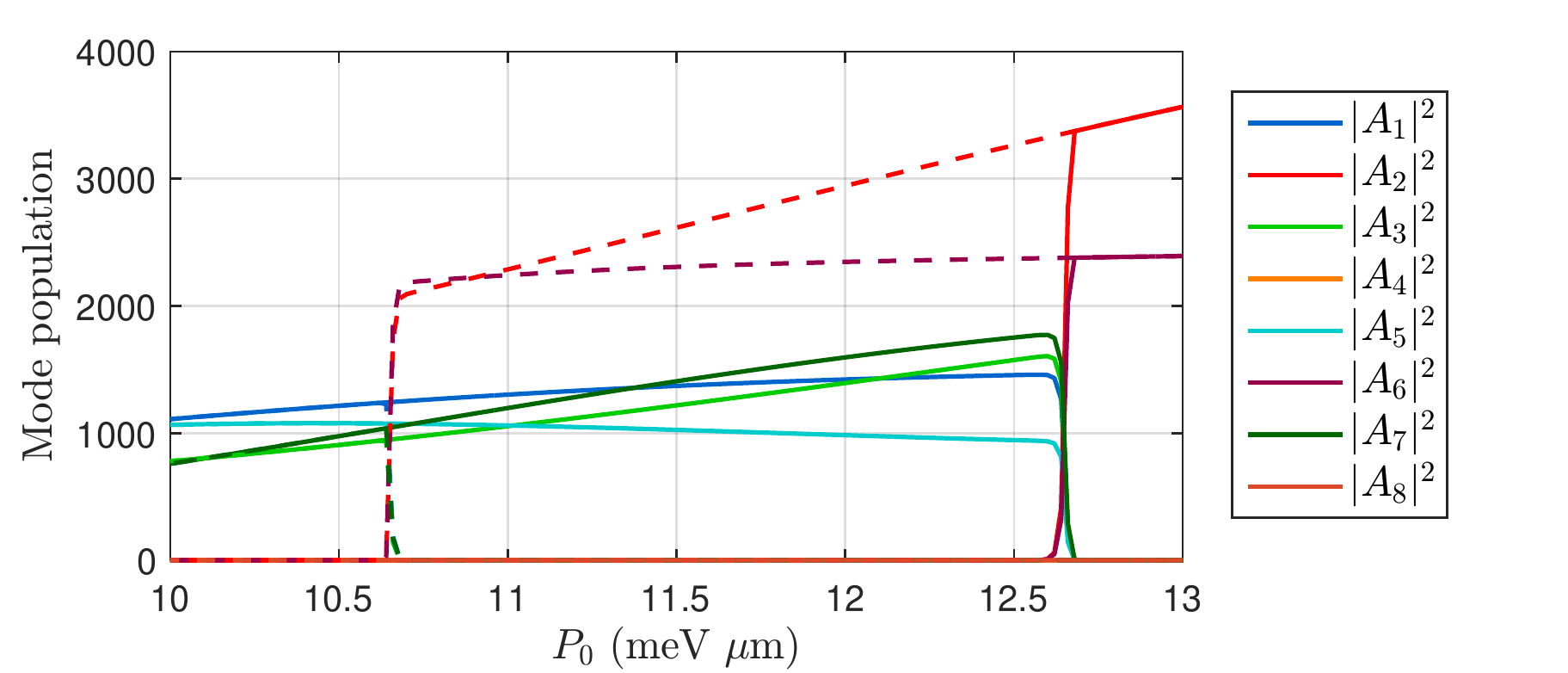} 
		\caption{(Color online) Bistable area demonstrated in the infinite quantum well of width $L = 60$ $\mu$m, the result is analogous to the bistable area displayed in Fig.~\ref{fig.res1D}. Parameters: $\alpha = 2.4$ $\mu$eV $\mu$m, $R = 0.3 \alpha$, $\tau = 2$ ps, $m = 3 \cdot 10^{-5} m_0$.}
\label{fig.res1D_IQW}
\end{figure}

Another example of a well known trapping geometry is the infinite quantum well, a well studied system with known eigensolutions to the Schrödinger equation. The main difference here from the HO is that the energy levels are no longer equidistant but grow quadratically with the mode quantum number $n$. Applying the same method as in Sec.~\ref{sec.cGP}[a] we see that a bistable interval also exists (see Fig.~\ref{fig.res1D_IQW}). Note the indexing of the modes in Fig.~\ref{fig.res1D_IQW} follows common literature on the infinite quantum well, where the quantum number starts at $n=1$ whereas in the HO it starts at $n = 0$.

\section{Conclusions}
We have shown that an open-dissipative polariton condensate, supported by incoherent pumping, described by a Gross-Pitaevskii type equation confined in a trapping potential can possess bistable regions. The bistability is inherited from firstly scattering between degenerate condensate modes, and secondly gain-decay mechanisms resulting from the different pump overlap for different harmonic modes. Results show that the minimum requirement for such bistability to take place is a system containing only two separate energy levels.

For higher pumping powers the condensate undergoes a series of oscillations in parity between highly occupied odd and even modes. We can switch from one solution of the condensate to its bistable counterpart in 2D harmonic potential guides, carrying information at velocities around $\sim 1.2$ $\mu$m/ps, and managing to travel along curved guides.

\section{Acknowledgements}
H. S. and I. S. acknowledge support by the Rannis Project BOFEHYSS and FP7 ITN project NOTEDEV.

\appendix
\begin{widetext}
\section{Two-Mode Model} \label{sec.app1}
Starting from Eq.~\ref{eq.ham2}, we come to the following two coupled equations for the first two eigenstates of the harmonic oscillator (neglecting higher modes) described by the cGP equation.
 \begin{equation}
i \hbar \frac{d A_0}{dt}  = \left(E_0- \frac{i\hbar}{2 \tau}\right) A_0 + i P_0 M_{0011} A_0  + (\alpha - i R ) \left[ M_{0000}|A_0|^2 A_0  +   M_{0011} \left(A_0^* A_1^2 + 2 A_0 |A_1|^2 \right) \right]
\label{eq.mode21}
\end{equation} 
\begin{equation} 
 i \hbar \frac{dA_1}{dt} = \left(E_1 - \frac{i\hbar}{2 \tau}\right) A_1 +i P_0 M_{1111} A_1 
 +  (\alpha - i R ) \left[  M_{1111}|A_1|^2 A_1 + M_{0011} \left( 2 |A_0|^2 A_1 +A_0^2 A_1^* \right)  \right]. \label{eq.mode22}
\end{equation}
Let's keep in mind that the matrix elements can be written in a more simpler form,
\[M_{0000} = \sqrt{ \frac{m \omega}{2 \pi \hbar} } =  \beta, \qquad M_{0011} = \frac{\beta}{2}, \qquad M_{1111} = \frac{3\beta}{4}. \]

\subsection{Noninteracting Case}
Though the coupled dynamical equations offer four independent equations to solve the problem exactly for any finite population in both modes, the nonlinear nature of the coupling does not reveal a clear solution depending on the physical parameters of the problem. It remains then unproven whether for some set of $(\omega, P_0, R)$ one can have solution with finite populations in both modes in the absence of interactions. 

However, one can determine two trivial stationary solutions: $|A_0|^2 = 0$, $|A_1|^2 = P_0/R$, and $|A_0|^2 = P_0/2R$, $|A_1|^2 = 0$. In princible, one can then have bistability if both solutions are stable. To check this, we perform stability analysis analogous to Ref.~\onlinecite{sarchi_2008} on the two solutions. We use the following ansatz for the former solution,
\[A_0(t) = e^{-i \mu t} \left( A_0^0 + \delta A_0(t) \right), \qquad A_1(t) = e^{-i \mu t} \left(0 + \delta A_1(t) \right) \]
Let's plug into Eqs.~\ref{eq.mode21}-\ref{eq.mode22} and neglect higher order terms of $\delta A_{0,1}$. The steady state will satisfy $|A_0^0| = P_0/2R$ and $\hbar \mu = E_0$.
\begin{equation} 
i \hbar \frac{d(\delta A_0)}{dt}  =   -i P_0  \frac{\beta}{2}  \delta A_0  - i  P_0 \beta \frac{(\delta A_0)^*}{2},
\end{equation} 
\begin{equation} 
 i \hbar \frac{d(\delta A_1)}{dt} = \left( \hbar \omega  + i P_0 \frac{\beta}{4}  \right) \delta A_1  - i P_0 \frac{\beta}{2} \frac{ (\delta A_1)^*}{2} . 
\end{equation}
The fluctuations can be described with $\delta A_{0,1}(t) = u_{0,1} e^{i \varepsilon t } + v_{0,1}^* e^{-i \varepsilon^* t}$ where $\hbar \varepsilon$ is the fluctuation energy. We then come to an eigenvalue problem $\mathcal{M} \delta \Psi = \varepsilon \delta \Psi$ where $\delta \Psi := (u_0, v_0, u_1, v_1)^T$. The matrix is,
\begin{equation}
\mathcal{M} = 
\begin{pmatrix}   -  iP_0 \beta/2  & -i P_0\beta/2 & 0 & 0 \\
-i P_0 \beta/2 &  -  iP_0 \beta/2  & 0 & 0 \\
0 & 0 & \hbar \omega  +  iP_0 \beta/4 & - iP_0 \beta/4  \\
0 & 0 & -iP_0 \beta/4 &  -\hbar \omega +  iP_0 \beta/4   \end{pmatrix}, \label{eq.a0mat}
\end{equation}
and for the case of $A_0(t) = e^{-i \mu t} \left(0 + \delta A_0(t) \right)$ and $A_1(t) = e^{-i \mu t} \left(A_1^0 +  \delta A_1(t) \right)$ we have,
\begin{equation}
\mathcal{M} = 
\begin{pmatrix}   -\hbar \omega  -  iP_0 \beta/2  & -i P_0\beta/4 & 0 & 0 \\
-i P_0 \beta/4 & \hbar \omega  -  iP_0 \beta/2  & 0 & 0 \\
0 & 0 & -  i3P_0 \beta/4  & - i3P_0 \beta/8  \\
0 & 0 & -i3P_0 \beta/8 &  -  i3P_0 \beta/4    \end{pmatrix}. \label{eq.a1mat}
\end{equation}
The spectrum will then consist of four eigenvalus $\varepsilon_i$ corresponding to the normal modes of $\delta A_{0,1}$. Here the stability is determined by the imaginary part of the eigenvalues. If all imaginary parts are negative the fluctuations decay exponentially. If however one or more eigenvalue is positive the the fluctuations grow exponentially.  In Fig.~\ref{fig.stab} we plot the imaginary parts of eigenvalues for Eqs.~\ref{eq.a0mat}-\ref{eq.a1mat} for different values of $P_0$ and $\omega$. We find that positive imaginary parts of the eigenvalues exist when $A_0$ is populated underlining that it is unstable whereas when we have only population in $A_1$ it is stable.
\begin{figure}[h!]
\centering
\includegraphics[width=1\textwidth]{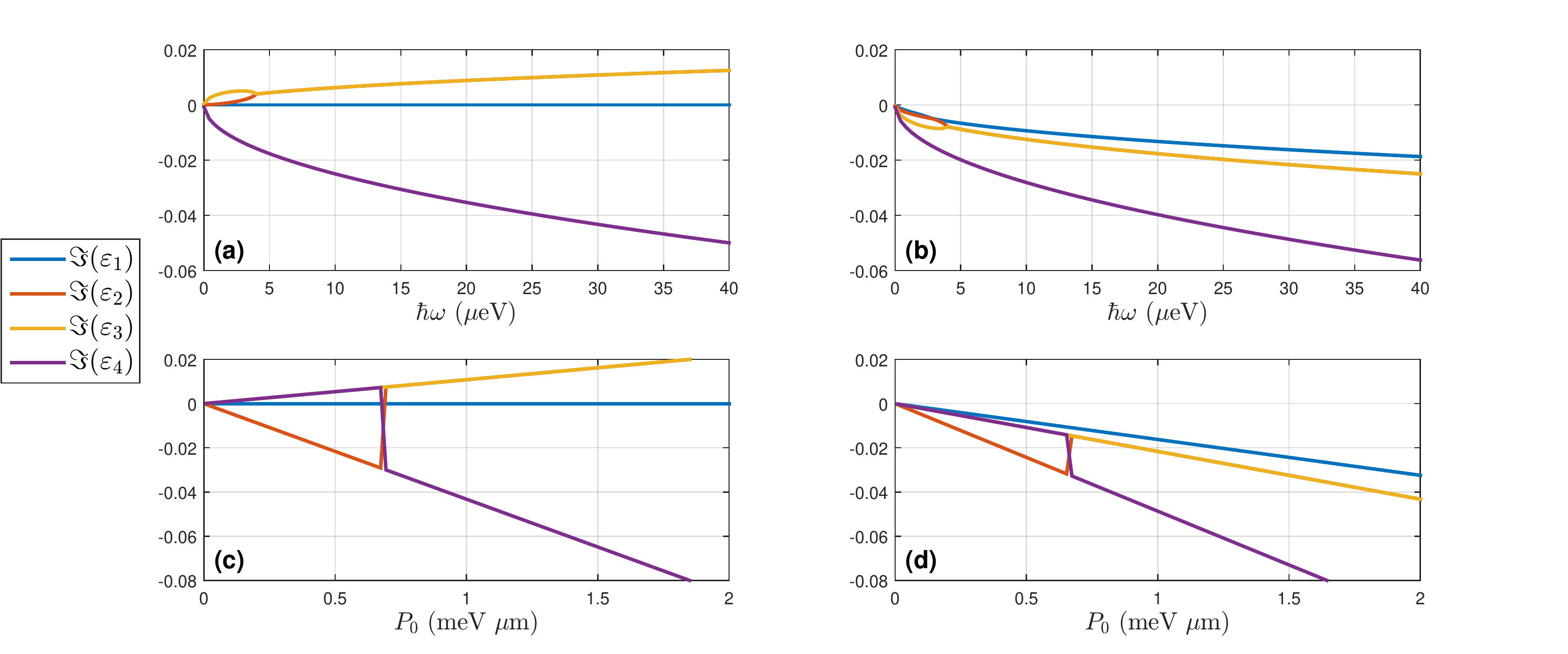}
\caption{(Color online) (a-b) Imaginary parts of the eigenvalues given by Eq.~\ref{eq.a0mat}-\ref{eq.a1mat} respectively for varying oscillator frequency $\omega$ and $P_0 = 1$ meV $\mu$m. (c-d) Same for $\hbar \omega = 30$ $\mu$eV and varying pump intensity $P_0$. For all plots: $R = 0.72$ $\mu$eV $\mu$m.}
\label{fig.stab}
\end{figure}

\subsection{Interacting Case}
In order to find an analytical expression for the two modes when they are both populated and interactions are nonzero, we make an attempt at solving Eqs.~\ref{eq.mode21} and~\ref{eq.mode22} for some arbitrary complex solution $A_0 = (a+ib) e^{-i \mu t}$ and $A_1 = c e^{-i \mu t}$ where we have assumed that the two modes are degenerate with condensate energy $\hbar \mu$ and the global phase is invariant (only relative phase is important). We then arrive at,
 \begin{align}
 \hbar \mu &  = \frac{\hbar \omega}{2} - \frac{i \hbar}{2 \tau} + (\alpha - i R ) \Big[
 M_{0000}(a^2 + b^2)  +  M_{0011} c^2 \frac{a^2 - b^2 - 2i ab}{a^2 + b^2} + 2M_{0011} c^2  \Big] + i P_0 M_{0011} \\
\hbar \mu & = \frac{3 \hbar \omega}{2} - \frac{i \hbar}{2 \tau} + (\alpha - i R ) \Big[
 M_{1111} c^2  +  M_{0011} (a^2-b^2 + 2iab)  + 2M_{0011} (a^2+b^2)  \Big]  + i P_0 M_{1111}
\end{align}
 From the first equation:
 \begin{align} \label{eq.re1}
 0 = & \left\{ \frac{\hbar \omega}{2} - \hbar \mu + \alpha \beta (a^2 + b^2) + \alpha \beta c^2 \frac{a^2 - b^2}{2(a^2 + b^2)} + \alpha \beta c^2 -  R \beta \frac{ab c^2}{a^2 + b^2} \right\}_{\text{Re}} \\
\label{eq.im1}  & - i\left\{  \frac{\hbar}{2 \tau_P} + \alpha \beta c^2   \frac{ab}{a^2 + b^2} + R \beta (a^2 + b^2) + R \beta c^2 \frac{a^2 - b^2}{2(a^2 + b^2)} + R \beta c^2 - P_0 \frac{\beta}{2}\right\}_{\text{Im}} 
\end{align}
From the second equation:
 \begin{align} \label{eq.re2}
0 = & \left\{ \frac{3 \hbar \omega}{2} - \hbar \mu + \alpha \frac{3 \beta c^2}{4} + \alpha \frac{\beta}{2} (a^2 - b^2)  + \alpha \beta (a^2 + b^2) +  R \beta ab \right\}_{\text{Re}} \\
\label{eq.im2} & - i\left\{  \frac{\hbar}{2 \tau_P} - \alpha \beta  ab + R \frac{3 \beta}{4} c^2 + R \beta  \frac{a^2 - b^2}{2} + R \beta (a^2 + b^2) - P_0 \frac{3 \beta}{4} \right\}_{\text{Im}} 
\end{align}
The real and imaginary parts must be zero. Let's take the difference of the real parts and imaginary parts respectively to get two separate equations for $c^2$ and to get rid of $\mu$ and $\tau$ for the time being.
\begin{align} \label{eq.cRe}
c^2 & = \frac{2\alpha  (a^4 - b^4) + 4R  a b(a^2 + b^2) + 4(\hbar \omega/\beta)(a^2 + b^2)}{2 \alpha  (a^2 - b^2) + \alpha (a^2 + b^2)  -  4R ab}, \\
 c^2 & = \frac{2 R  (a^4 - b^4) - 4 \alpha  ab (a^2 + b^2) - P_0  (a^2 + b^2)}{2R  (a^2 - b^2) + R  (a^2 + b^2) + 4 \alpha  ab}. \label{eq.cIm}
\end{align}
Equating the two, we arrive at a cubic equation for $b$,
 \begin{equation}
0 = -b^3 \left[3a(\alpha^2 + R^2) \right] - b^2  \left[ \frac{\hbar \omega}{\beta} R  + \frac{P_0}{4} \alpha \right] + b \left[ 5a^3(\alpha^2 + R^2) + \frac{4\hbar \omega}{\beta} \alpha a  - P_0R a \right] + \left[ \frac{3 \hbar \omega}{\beta} R a^2 +  \frac{3 P_0}{4} \alpha a^2  \right]. \label{eq.cube}
\end{equation}
Thus, the bistable area corresponds to an interval where three real roots exist to this equation for a real valued $c$ and satisfy the constraint that Eqs.~\ref{eq.re1}-\ref{eq.im2} should be zero. This allows us to solve the hysteresis branches of bistable areas as shown in Fig.~\ref{fig.2mode} (green and black dots). 

\section{Three-Mode Model} \label{sec.app2}
Working with only the first three eigenstates of the HO. One arrives at three coupled equations,
 \begin{align} \notag
& i \hbar \frac{dA_0}{dt}  = \left(E_0 - \frac{i\hbar}{2\tau}\right) A_0 + i P_0 ( M_{0011} A_0 + M_{0112} A_2 ) + (\alpha - i R ) \bigg\{ M_{0000}|A_0|^2 A_0 +  M_{0222}|A_2|^2 A_2  \dots  \notag \\
&  +  M_{0002}\Big[2|A_0|^2 A_2+ A_0^2 A_2^*\Big]  +  M_{0011} \Big[A_0^* A_1^2 + 1 A_0 |A_1|^2 \Big] +  M_{0022}\Big[A_0^* A_2^2 + 2 A_0 |A_2|^2\Big] +  M_{0112}\Big[ 2|A_1|^2 A_2  + A_1^2 A_2^*\Big] \bigg\}, \label{eq.3mode1}
\end{align} 
\begin{align} \notag
& i \hbar \frac{dA_1}{dt}  = \left(E_1 - \frac{i\hbar}{2\tau}\right) A_1 +i P_0 M_{1111} A_1 +  (\alpha - i R ) \bigg\{  M_{1111}|A_1|^2 A_1 \dots \notag \\ 
& + M_{0011}\Big[ 2 |A_0|^2 A_1 +A_0^2 A_1^* \Big] +  2 M_{0112} \Big[ A_0^* A_1 A_2 + A_0 A_1^* A_2 +  A_0 A_1 A_2^* \Big] + M_{1122} \Big[ A_1^* A_2^2 + 2A_1 |A_2|^2 \Big]   \bigg\}, \label{eq.3mode2}
\end{align}
\begin{align} \notag
& i \hbar \frac{dA_2}{dt}  = \left(E_2 - \frac{i\hbar}{2 \tau}\right) A_2 +  i P_0 ( M_{0112} A_0 +  M_{1122} A_2 )  + (\alpha - i R ) \bigg\{  M_{2222}|A_2|^2 A_2 + M_{0002}|A_0|^2 A_0  \dots \notag \\ 
&   +  M_{0022} \Big[ 2|A_0|^2 A_2 + A_0^2 A_2^* \Big] + M_{0112} \Big[ A_0^* A_1^2 + 2 A_0 |A_1|^2 \Big] +   M_{0222} \Big[ A_0^* A_2^2 + 2  A_0 |A_2|^2 \Big] +  M_{1122} \Big[ 2|A_1|^2 A_2 + A_1^2 A_2^*\Big] \bigg\}. \label{eq.3mode3}
\end{align}
\end{widetext}

\bibliography{bibliography}

\end{document}